\documentclass[superscriptaddress,prx,longbibliography,twocolumn, footinbib]{revtex4-2}

\usepackage{subcaption}
\usepackage[space]{grffile}
\usepackage{amsmath, amssymb, amsfonts,bm}
\usepackage{latexsym}   
\usepackage{bbm} 
\usepackage{mathtools}  
\usepackage{placeins}   
\newcommand{\expe}[1]{\left\langle #1 \right\rangle}

\newcommand{\Harvard}{Department of Physics, Harvard University, Cambridge, Massachusetts 02138, USA.}

\newcommand{\ETH}{Institute for Theoretical Physics, ETH Zurich, 8093 Zurich, Switzerland.}

\newcommand{\Tech}{Department of Physics, Technion, 32000 Haifa, Israel}

\newcommand{\Calif}{Department of Physics, University of California San Diego, La Jolla, California 92093, USA}

\newcommand{\Hamb}{Max Planck Institute for the Structure and Dynamics of Matter, Luruper Chausse 149, 22761 Hamburg, Germany}

\newcommand{\Oxford}{Department of Physics, University of Oxford, Clarendon Laboratory, Parks Road, Oxford OX1 3PU, UK}

\begin{document}

\title{
Generalized Fresnel-Floquet equations for driven quantum materials
}

\author{Marios~H.~Michael}\email[Correspondence to: ]{marios\_michael@g.harvard.edu}
\affiliation{\Harvard}
\author{Michael F\"{o}rst}
\affiliation{\Hamb}
\author{Daniele Nicoletti}
\affiliation{\Hamb}
\author{Sheikh~Rubaiat~Ul~Haque}
\affiliation{\Calif}
\author{Andrea Cavalleri}
\affiliation{\Hamb}
\affiliation{\Oxford}
\author{Richard~D.~Averitt}
\affiliation{\Calif}
\author{Daniel~Podolsky}
\affiliation{\Tech}
\author{Eugene~Demler}
\affiliation{\Harvard}
\affiliation{\ETH}

\date{\today}

\begin{abstract}
Optical drives at terahertz and mid-infrared frequencies in quantum materials are increasingly used to reveal the nonlinear dynamics of collective modes in correlated many-body systems and their interplay with electromagnetic waves. Recent experiments demonstrated several surprising optical properties of transient states induced by driving, including the appearance of photo-induced edges in the reflectivity in cuprate superconductors, observed both below and above the equilibrium transition temperature. Furthermore, in other driven materials, reflection coefficients larger than unity have been observed. In this paper we demonstrate that unusual optical properties of photoexcited systems can be understood from the perspective of a Floquet system; a system with periodically modulated system parameters originating from pump-induced oscillations of a collective mode. We present a general phenomenological model of reflectivity from Floquet materials, which takes into account parametric generation of excitation pairs. We find a universal phase diagram of drive induced features in reflectivity which evidence a competition between driving and dissipation. To illustrate our general analysis we apply our formalism to two concrete examples motivated by recent experiments: a single plasmon band, which describes Josephson plasmons in layered superconductors, and a phonon-polariton system, which describes upper and lower polaritons in materials such as insulating SiC. Finally we demonstrate that our model can be used to provide an accurate fit to results of phonon-pump - terahertz-probe experiments in the high temperature superconductor $\rm{YBa_2Cu_3O_{6.5}}$. Our model explains the appearance of a pump-induced edge, which is higher in energy than the equilibrium Josephson plasmon edge,  even if the interlayer Josephson coupling is suppressed by the pump pulse. 
\end{abstract}

\maketitle


\section{Introduction and overview}

\subsection{Motivation}
Nonequilibrium dynamics in quantum materials is a rapidly developing area of research that lies at the interface between nonlinear optics and quantum many-body physics\cite{Aoki2014,delatorre2021nonthermal}. Indeed, a panoply of experimental results highlight the ability of ultrafast optical techniques to interrogate and manipulate emergent states in quantum materials. This includes photo-augmented superconductivity \cite{Fausti2011,Nicoletti2014,Hu2014,Okamoto2016}, unveiling hidden states in materials proximal to the boundary of an insulator-to-metal transition
\cite{Zhang2016}, and manipulating topological states \cite{Wang2013,Sie2019,McIver2019}. The
terahertz to mid-infrared spectral range is especially important as numerous phenomena in quantum materials manifest at these energies, including phonons, magnons, plasmons, and correlation gaps\cite{Basov2011}. Access to this spectral range enables preferential pump excitation of specific degrees of freedom and probing of the resultant dynamics that are encoded in the dielectric response (and hence
the reflectivity or transmission). Therefore, a particular challenge is to decode the optical reflectivity dynamics which typically requires developing models that can be related to the underlying microscopic states. In short, it is crucial to develop a consistent framework for interpreting experimental results to aid
in identifying emergent universal properties of driven states and to take full advantage of the plethora of “properties-on-demand” exhibited by quantum materials \cite{Basov2017}.

So far, the predominant paradigm of understanding pump and probe experiments has been based on the
perspective of a dynamic trajectory in a complex energy landscape, where “snapshots” track the evolution of the slowly evolving but quasi-stationary many-body states \cite{Sun2020}. Within this approach temporal evolution of spectroscopic features is interpreted using the conventional equilibrium formalism, and measured parameters serve as a fingerprint of the underlying instantaneous state. In particular, this approach has been applied to analyze c-axis terahertz reflectivity of the cuprate superconductors. In equilibrium, the Josephson plasma (JP) edge appears only below T$_{c}$ indicating coherent c-axis Cooper-pair tunneling. Interband or phononic excitation along the c-axis in several distinct cuprates (including La$_{1.675}$Eu$_{0.2}$Sr$_{0.125}$CuO$_{4}$, La$_{2-x}$Ba$_{x}$CuO$_{4}$, YBa$_{2}$Cu$_{3}$O$_{6+\delta}$) resulted in the appearance of edge-like features in the terahertz c-axis reflectivity at temperatures above the equilibrium T$_{c}$\cite{Fausti2011,Nicoletti2014,Hu2014,Cremin2019}. These experiments were interpreted as providing spectroscopic evidence for light induced modification of interlayer Josephson coupling. The central goal
of this paper is to develop an alternative framework for interpreting optical responses of photoexcited materials. Our model focuses on features unique to nonequilibrium systems, in particular to photoexcited collective excitations which provide parametric driving of many-body systems. While we do not argue that this mechanism explains all experimental results on pump induced changes in reflectivity, we believe
that this scenario is sufficiently ubiquitous to merit detailed consideration. We provide universal
expressions for driving induced changes in the reflectivity, which can be compared to experimental data, in order to examine the relevance of the parametric driving scenario to any specific experiment.

Before proceeding to discuss details of our model, it is worth reviewing several experiments that have already revealed pump-induced dynamics that cannot be interpreted from the perspective of equilibrium systems. Particularly striking are recent observations of light amplification in the photoexcited insulator SiC and the superconductor K$_{3}$C$_{60}$ above its equilibrium T$_{c}$ \cite{Cartella2018, budden20,Buzzi21}. Furthermore, in the case of pumped YBa$_{2}$Cu$_{3}$O$_{6+\delta}$ discussed above, strong experimental evidence has accumulated indicating that an effective photo-induced edge arises from parametric amplification of Josephson plasmons rather than a modification of the effective coupling \cite{vonHoegen19,Marios20} (see discussion in Section-\ref{sec:exp} of this paper). Prior work also demonstrated higher harmonic generation from Higgs and other collective modes\cite{Shimano2020,Giorgianni2019,Gabriele2021} and nonlinear effects including parametric amplification of superconducting plasma waves\cite{Denny2015,Rajasekaran2016,Dienst2013,Rajasekaran2018}. A cursory understanding of these
experiments can be obtained from the perspective of nonlinear optics deriving from coherent dynamics of order parameters and associated degrees of freedom such as phonons. However, several qualitative differences between collective mode optics and standard nonlinear optics deserve a special mention. First, in systems that we consider, a nonlinear response of the probe pulse persists at delay times well beyond
the duration of the pump pulse. Hence, one cannot apply theoretical approaches based on the expansion of optical nonlinearities in powers of the electric field, such as $\chi^{(2)}$, and $\chi^{(3)}$ \cite{Boyd2008}. Instead, it is imperative to analyze the interaction of the electromagnetic field of the probe pulse with
matter degrees of freedom excited by the pump pulse. Second, it is important to properly account for the role of the surface, since the probe wavelength can be comparable or even larger than the penetration depth of the material. Thus, common assumptions of nonlinear optics, including the slowly varying envelope approximation \cite{Boyd2008} and phase matching, need to be replaced by the explicit solution of Maxwell equations coupled to dynamical equations for matter. 
\begin{figure}[t!]
    \centering
    \includegraphics[trim= 1.6cm 1.5cm 1cm 1cm, clip, scale = 0.45]{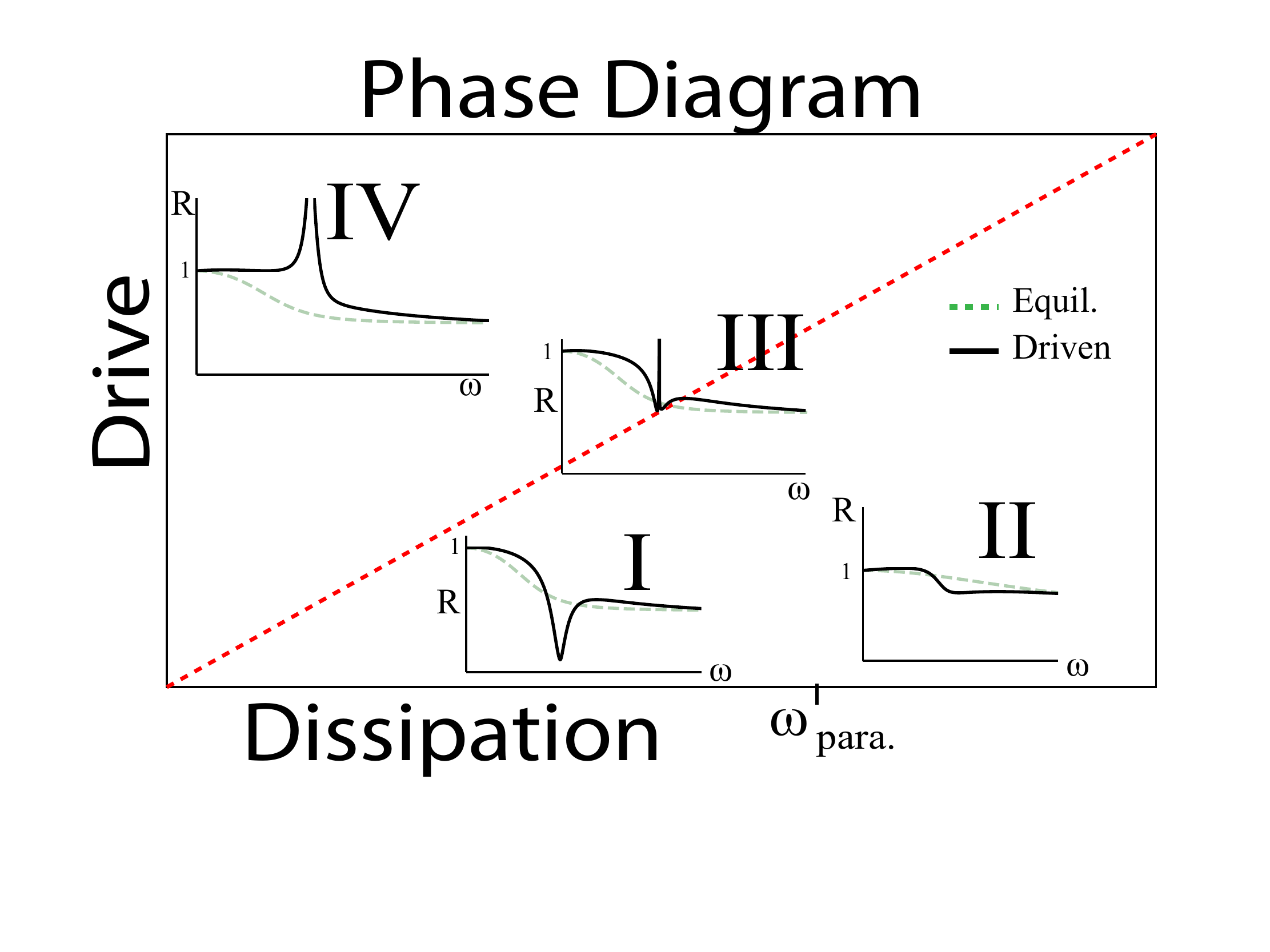}
    \caption{Phase diagram of optical reflectivity of an interacting Floquet material as a function of the parametric drive amplitude and dissipation.
    We identify four regimes with qualitatively different types of reflectivity: (I) Weakly driven underdamped modes in the stable regime where dissipation, $\gamma$, is sufficient to prevent a lasing instability. The line shape is a square lorentzian dip given by equation~(\ref{eq:RegimeI}).  (II) Weakly driven overdamped modes in the stable regime. The resonance feature is an edge-like line shape given by equation~(\ref{eq:overdamped}). (III) Cross-over regime on the boundary of the stable and unstable regions with a double dip structure. (IV) Unstable region, strong driving overcomes dissipation and may even lead to parametric amplification.  }
    \label{fig:phase}
\end{figure}

\subsection{Overview of results and organization of the paper}
\label{SectionOverview}

The primary goal of this paper is to present a general phenomenological formalism for discussing optical properties of driven states following a resonant excitation of a collective mode. We analyze the problem from the perspective of Floquet matter, in which a collective mode excited by the pump provides temporal modulation of microscopic parameters of the material. This results in parametric driving of the system and Floquet-type mixing of frequencies. When the system is driven homogeneously in space (i.e. with wave vector $k=0$) and frequency $\Omega_d$, a parametric resonance occurs whenever two collective excitations that are IR-active have the opposite wave vector,  $k_1=-k_2$, and frequencies that add up to to the drive frequency, $\omega_1+\omega_2=\Omega_d$. Naively, one expects parametric resonances to always lead to enhancement of reflectivity, with sharp peaks corresponding to parametric resonance conditions. We find that the situation is far richer and may include the appearance of edges, dips, and electromagnetically induced transparency (EIT)\cite{Harris96} type structure in the reflectivity (see Fig.~\ref{fig:phase}). Physically, this comes from oscillation induced mixing between light-matter fields of different frequency components. In this paper, we focus on the case of oscillations with a small amplitude and/or strong dissipation, in which case analysis can be limited to the mixing of only two frequencies, commonly referred to as the signal and idler frequencies. They are defined such that the signal frequency $\omega_s$ is equal to the frequency of the probe pulse, whereas the idler frequency $\omega_{\rm id.}$ is equal to the difference between the drive frequency $\Omega_d$ and $\omega_s$. Interference between the signal and idler frequency components is reminiscent of Fano-type phenomena in scattering theory and optics, where interference between two channels can result in non-trivial energy dependence of the reflection and transmission coefficients\cite{Limonov17}. Describing the driving only in terms of signal and idler mixing corresponds to a degenerate perturbation theory in the Floquet basis \cite{Eckardt15,Sho19,Buzzi21}. 

What determines whether interference phenomena will dominate over parametric amplification of reflectivity is the competition between parametric driving and losses. We find a universal dynamical phase diagram of the optical response as a function of the strength of the drive and dissipation. Remarkably, we find that the entire breadth of these responses can be specified using only a few effective (phenomenological) parameters. One of the main achievements of this paper is to derive analytical formulas for the shape of these resonances in section~\ref{sec:phaseDiag} in the case of strong dissipation where perturbation theory is valid, Regimes I and II in Fig.~\ref{fig:phase}. In Regime I, corresponding to the case of underdamped collective modes, we obtain a Lorentzian square shape:
\begin{equation}
    R_{\rm driven} = R_s \left( 1 + \alpha \mbox{Re} \{ \frac{1}{\left(\omega - \omega_{\rm para.} + i \gamma \right)^2} \} \right),
\label{eq:RegimeI}
\end{equation}
where $R_{\rm driven}$ is the reflectivity in the Floquet state, $R_s$ the reflectivity in equilibrium, $\alpha$ a frequency dependent parameter that depends on dispersion of the IR collective modes of the material, $\omega_{\rm para.} $ the frequency at which parametric resonance condition is satisfied and $\gamma$ the dissipation in the system. Notably, in Regime I, we can use the Floquet drive to directly extract the dissipation in the system on parametric resonance. In Regime II, corresponding to overdamped collective modes, the resonance peak has the form:
\begin{equation}
     R_{\rm driven} = R_s \left( 1 + \beta \mbox{Re} \{ e^{i \theta} \frac{1}{\omega - \omega_{\rm para.} + i \gamma } \} \right),
\label{eq:RegimeII}
\end{equation}
where $\beta$ and $\theta$ are frequency dependent parameters that depend on the dispersion of IR collective modes. In this case, the shape is a linear combination of a real and imaginary lorentzian function resulting in an effective "edge" like feature. 

For clarity, in this work we simplify our analysis by including Floquet modulation at a single frequency. When the finite lifetime of the collective mode is taken into account, this should be analyzed as a multi-tonal drive. Our analysis can be generalized to this situation. However, in the current paper we will only comment on the main changes that we expect in this case. We postpone a detailed discussion of the multi-tonal Floquet-Fresnel problem to a future publication \cite{Dolgirev_unpublished}. 

It is worth noting conceptual connections between our Floquet approach and previous work on the phenomenon of optical phase conjugation (OPC) \cite{Fisher1983,ZelDovich85}. What makes our analysis different is that we focus on terahertz phenomena, which correspond to much longer wavelengths than optical phenomena considered in the context of OPC. It is important for our discussion to take into account that non-linear processes take place near the material boundary rather than in the bulk, which is why our starting point is the Fresnel formalism of reflection of electromagnetic waves. This can be contrasted to phase matching conditions used in most discussions of OPC, which essentially assume that non-linear processes take place in the bulk of the material.

\begin{figure}
    \centering
    \includegraphics[trim= 6cm 3cm 2cm 3cm, clip, scale = 0.70]{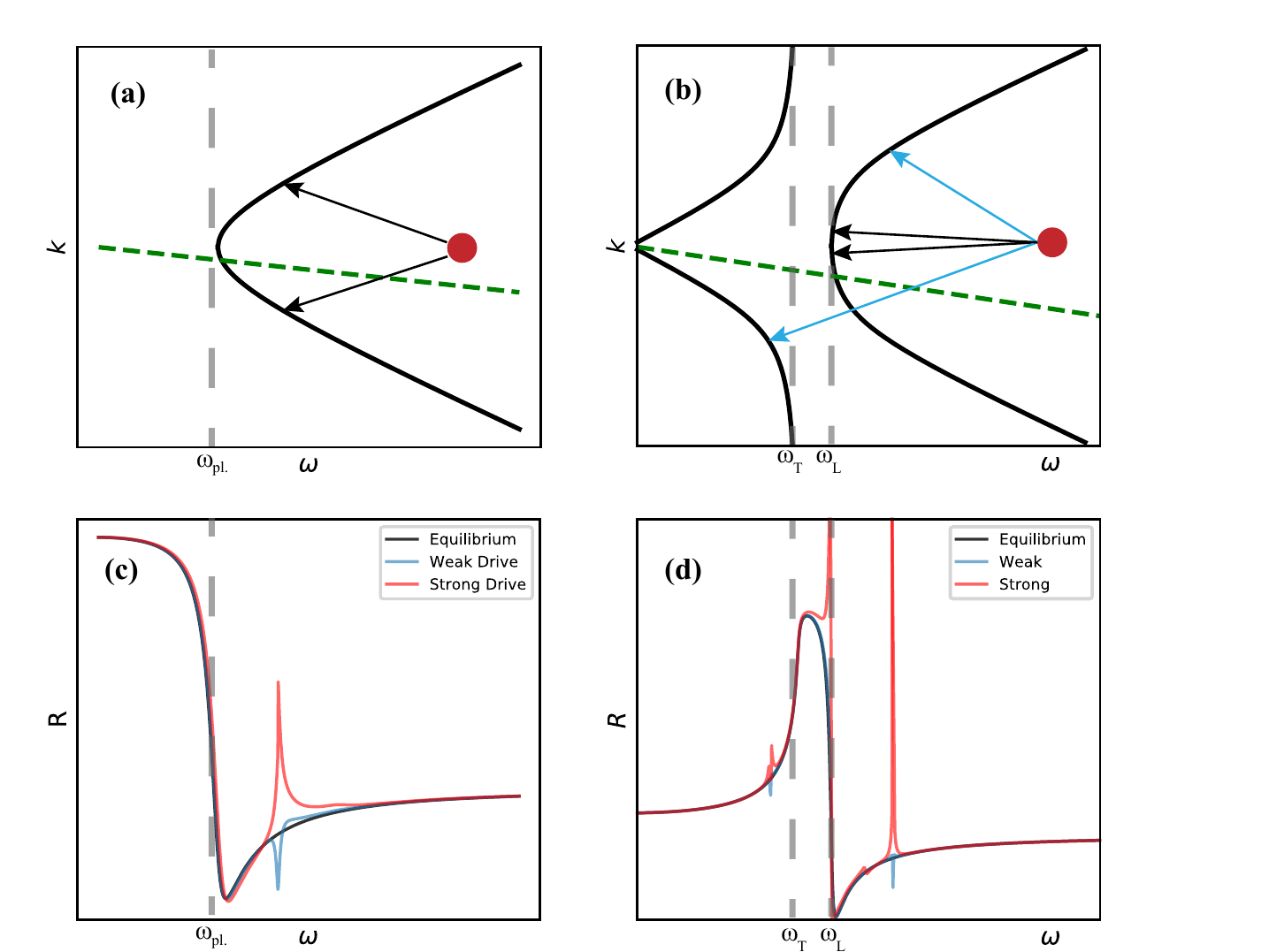}
    \caption{Examples of dispersion relations (a) - (b) and their corresponding reflectivity spectrum (c) - (d). a) Dispersion relation of a plasmon in a SC (black) and dispersion relation of light in air (green). The corresponding equilibrium reflectivity in (c) shows perfect reflection below the gap, for $\omega_s < \omega_{\rm pl.}$, and a minimum in reflectivity appears when the dispersion of light in air crosses the dispersion of the plasmon in the material, a condition called phase matching. (b) Dispersion relation of phonon-polariton (black) and dispersion relation of light in air (green). The corresponding equilibrium reflectivity in (d) shows perfect reflectivity inside the \textit{reststrahlen band} and plasma edge when the dispersion of light in air crosses the dispersion of the upper polariton, similarly to (c). The red dot in (a) and (b) show the driving frequency, while the arrows depict the parametrically resonant processes resulting from the Floquet drive. This leads to features in the reflectivity predicted by our theory at the parametrically resonant frequencies both for strong and weak drive relative to dissipation. }
\label{fig:equil}
\end{figure}

This paper is organized as follows. Section ~\ref{SectionFormalism} presents a general formalism for computing the reflectivity of Floquet materials. With a goal of setting up notation in section \ref{sec:equil} we remind the readers the canonical formalism of Fresnel's solution of light reflection from an equilibrium material with an index of refraction $n(\omega)$. In Section~\ref{sec:floq} we discuss how to generalize this approach to study light reflection from a material subject to a periodic drive. We show a universal form of frequency dependence of reflectivity from such systems, which we  summarize in the phase diagram presented in Figure~\ref{fig:phase}. We show that this frequency dependence can be deduced from the dispersion of collective modes and frequency of the periodic drive without developing a full microscopic theory. Thus the Floquet-Fresnel equations allows for the same level of conceptual understanding as the standard equilibrium Fresnel problem. To make our discussion more concrete, in section \ref{sec:parad} we apply this analysis to two  paradigmatic cases: i) a single low frequency plasmon band and ii) the two band case of a phonon-polariton system with dispersions shown in Figure~\ref{fig:equil}. These two cases are not only exemplary but also provide accurate models for real materials, such as the lower Josephson plasmon of $\rm{YBa_2Cu_3O_{6+x}}$ (case (i)) and a phonon-polariton system in $\rm SiC$ (case (ii)).They are reviewed in sections \ref{SectionPlasmaBand} and \ref{sec:SiC} respectively. We note that most cases of parametric resonance in pump and probe experiments can be reduced to these two examples, since the usual formulation of parametric resonance involves generating pairs of excitations and the resonance can be described by including up to two bands. However, in some cases there may be additional features in the reflectivity arising from singular behavior of matrix elements. We provide a concrete example of this in section~\ref{sec:SiC} for the case of phonon-polariton model in SiC. Finally, we demonstrate that our theory enables a quantitatively accurate fit to the  results of pump and probe experiments in $\rm{YBa_2Cu_3O_{6.5}}$. These experiments demonstrated the appearance of a photo-induced edge both below and above the superconducting transition temperature, at frequencies close to the lower plasmon edge. We demonstrate that these observations can be accurately described by the Floquet-Fresnel model developed in this paper. 

\section{General formalism of Floquet-Fresnel reflection} 
\label{SectionFormalism}

\subsection{Equilibrium reflectivity}
\label{sec:equil}
We begin our discussion by presenting coupled dynamical equations for light and matter, assuming that the material has an infrared active collective mode, such as a phonon or a Josephson plasmon. Information about the collective mode is included in the frequency dependence of the linear electric susceptibility,  $\chi(\omega,k)$, which determines the index of refraction $n(\omega)$:
\begin{subequations}
\begin{align}
 \nabla \times B =& \mu_0 \partial_t D ,\\
 \nabla \times E =& - \partial_t B , \\
 D =& \epsilon_0 E + P 
\end{align}
\end{subequations}
where the dynamics of the polarization $P$ contain all optically active collective modes inside the material, $E$ is the electric field and $\epsilon_0$ and $\mu_0$ are the electric permittivity and magnetic permeability in vacuum, respectively. The polarization in frequency and momentum space, $P(\omega, k) $, is given in terms of the electric field through the linear susceptibility, $P(\omega, k) = \epsilon_0 \chi(\omega, k) E(\omega,k)$. Due to the high speed of light, $c$,  considerable hybridization between the collective mode and light occurs only at very small momenta, $k \sim \frac{\omega}{c} $. As a result, for optical reflection problems we can take the susceptibility to be dispersionless, $\chi(\omega,k) \approx \chi(\omega, k = 0) \equiv \chi(\omega)$ to a good approximation. Combining the Maxwell equations with the susceptibility we find the dynamics of the electromagnetic transverse modes in frequency and momentum space to be given by a wave equation with a solely frequency dependent refractive index $n(\omega)$: 
\begin{equation}
    \left( \frac{ n^2(\omega) \omega^2  }{c^2} - k^2 \right) E(\omega, k)  = 0.
\label{eq:eomEquil}
\end{equation}
Collective mode dispersion relations are found as solutions to the equation $\left( k^2 - \frac{ \omega^2 n^2(\omega) }{c^2} \right) = 0$. The above description is very general and any dispersion relation inside the material can be captured by an appropriate choice of $n(\omega)$.

\subsubsection{The case of a Plasmon}
\label{SectionPlasmaBand_n}

In superconductors (SC) the Anderson-Higgs mechanism gives rise to the gap in the spectrum of transverse electromagnetic fields equal to the plasma frequency, see Fig.~\ref{fig:equil}(a). The plasmon excitation can be captured by a refractive index of the type\cite{Tinkham04}:
\begin{equation}
    n^2_{SC} ( \omega) = \epsilon_{\infty} \left( 1 - \frac{\omega^2_{\rm pl.}}{\omega^2 } \right),
\end{equation}
where $\omega_{\rm pl.} $ is the plasma frequency and $\epsilon_{\infty}$ a constant background permittivity. Such a refractive index when substituted in equation~(\ref{eq:eomEquil}), leads to the dispersion relation for the electromagnetic field inside a SC to be:
\begin{equation}
    \omega_{SC}^2(k) = \omega^2_{\rm pl.} + \frac{c^2}{\epsilon_{\infty}} k^2. 
\end{equation}

We note that plasmon modes can have very different frequencies depending on light polarization. In particular, in the case of layered systems,  such as $\rm Y Ba_2 Cu_3 O_{6 +x} $ superconductors, the plasma frequency is small for electric field polarization perpendicular to the layers. In layered metals one can also find low energy plasmon modes, although they typically have stronger damping than in superconductors.

\subsubsection{Phonon-polariton systems}

Another ubiquitous example is the  case of phonon-polaritons. In this paper we will primarily use $\rm SiC$ for illustration, which features an IR-active phonon at frequency close to 30 THz with a large effective charge\cite{Cartella2018}. Another related material that is currently under active investigation is  $\rm Ta_2 Ni Se_{5}$, which has an additional complication that multiple phonons need to be included in the analysis.   

In the case of a single IR phonon the dispersion relation of the phonon-polariton system is depicted in Fig.~\ref{fig:equil} (b). It can be captured by substituting in equation~(\ref{eq:eomEquil}) the refractive index\cite{Sho19}:
\begin{equation}
    n_{\rm phonon}^2( \omega) = \epsilon_{\infty} \left( 1 - \frac{\omega_{\rm pl.,phonon}^2}{\omega^2 + i \gamma \omega  - \omega^2_{\rm ph.}} \right),
\end{equation}
where $\omega_{\rm pl.,phonon}$ is the plasma frequency of the phonon mode, $\omega_{\rm ph.}$ the transverse phonon frequency and $\gamma$ a dissipative term for the phonon. 

\subsubsection{The case of multiple IR modes}

In the case when multiple IR-active collective modes need to be included in the analysis (phonons, plasmons, etc.), it is common to use the Lorentz model which parametrizes the contribution of each collective mode to the refractive index by a lorentzian \cite{WOOTEN72}:
\begin{equation}
    n^2(\omega) = \epsilon_{\infty} \left(1 - \sum_{i} \frac{\omega_{pl.}^2}{ \omega^2 + i \gamma_i \omega - \omega_{i}^2} \right),
\end{equation}
where $\omega_i$ is the bare frequency of the $i$th collective mode, $\omega_{pl.,i}$ the plasma frequency which characterises the strength of the coupling to light, $\gamma_i$ the dissipation and $\epsilon_{\infty}$ an effective static component to the permittivity arising from high energy modes not included in the sum. The above discussion, illustrates that equation~(\ref{eq:eomEquil}) is very general, and in any case an appropriate $n(\omega)$ can be chosen to capture the dispersion relation of optically active bands.


\subsubsection{Fresnel equations}

We begin by reviewing the Fresnel light reflection problem at the interface between air and material in equilibrium. While this is a textbook material, we present it here with the goal of establishing notations for subsequent discussion of the non-equilibrium case. We  consider an incoming beam with frequency  $\omega_{\rm s}$ at normal angle of incidence 
\begin{equation}
    E_{s} = E_0 e^{ i \frac{\omega_s}{c} z - i \omega_s t}
\end{equation}
where $z$ is the direction perpendicular to the interface and the interface lies at $z = 0$. The reflected and transmitted waves at the signal frequency are expressed through reflection and transmission coefficients, $E_r = r_s E_0e^{ - i \frac{\omega_s}{c} z - i \omega_s t} $ and $E_t = t_s E_0e^{ i k_s z - i \omega_s t}$. The momentum $k_s$ corresponds to the mode inside the material oscillating at $\omega_s$ and using equation~(\ref{eq:eomEquil}), is given by $k_s = \frac{\omega_s n(\omega_s)}{c}$. Matching the electric field across the boundary at $z=0$ gives rise to the boundary equation:
\begin{equation}
1 + r_s = t_s.
\end{equation}
For non-magnetic materials, the magnetic field is also conserved across the surface. Using the homogeneous Maxwell equation, $\partial_t B = - \nabla \times E$, we calculate the magnetic field in the two regions. Matching the two regions at $z = 0$ gives rise to the second boundary equation:
\begin{equation}
    1 - r_s = n(\omega_s) t_s.  
\end{equation}
Solving for the reflection coefficient, we find the standard expression for reflectivity in terms of the refractive index:
\begin{equation}
\begin{split}
    R_s = |r_s|^2 = \left| \frac{1 - n(\omega_s)}{1 + n(\omega_s)}\right|^2 = \frac{\left( 1 - n'\right)^2 + \left(n''\right)^2 }{\left( 1 +  n'\right)^2 + \left(n''\right)^2}
\end{split}
\label{eq:reflEquil}
\end{equation}
where $n'$ and $n''$ correspond to the real and imaginary part of the refractive index respectively. 

In equilibrium, reflectivity can be deduced, at least qualitatively, from the form of collective mode dispersion inside the material. This is depicted in Fig.~(\ref{fig:equil})(a) -(c) for a SC and in Fig.~(\ref{fig:equil})(b) - (d) for a photon-polariton system. In the case of a SC, at probing frequencies below the plasma gap $\omega_s < \omega_{pl.}$, no bulk modes exist to propagate the energy and $k_s$ is purely imaginary corresponding to an evanescent wave. In this situation, we have near perfect reflectivity. As soon as the probing frequency becomes larger than the plasma gap, transmission is allowed and reflectivity drops abruptly, reaching a minimum at the frequency where the light-cone crosses the plasma band. The minimum in reflectivity or equivalently the maximum in transmission occurs when the incoming and transmitted waves are "phase matched", a condition that is satisfied when the light cone in air crosses a new band inside the material, i.e. $n'(\omega_s) = 1$ in equation~(\ref{eq:reflEquil}). The sudden drop in reflectivity appearing whenever a new optically active band becomes available is called in the literature a "plasma edge". Similar reasoning can be used to determine qualitatively the reflectivity of a phonon-polariton system from its dispersion relation alone: At frequencies within the gap of the dispersion, $\omega_{\rm ph.} < \omega_s < \sqrt{ \omega_{\rm ph.}^2 + \omega_{\rm pl., phonon }^2} $, called the \textit{reststrahlen band}, only evanescent waves are allowed, and reflectivity is expected to be close to one. On the other hand for probing frequencies $\omega_s > \sqrt{ \omega_{\rm ph.}^2 + \omega_{\rm pl}^2} $, when the light cone crosses the upper polariton branch, a plasma edge appears.



\subsubsection{Dissipation}

Finally, we comment on the effects of dissipation on light reflection. While in principle, equation~(\ref{eq:eomEquil}) is completely general, it is sometimes helpful to add dissipation explicitly through the conductivity of the normal electron fluid which obeys Ohm's law and modifies equation~(\ref{eq:eomEquil}) to:
\begin{equation}
    \left( n^2(\omega) \omega^2 + i \frac{\sigma_{n}}{\epsilon_0} \omega  - c^2 k^2 \right)E = 0,
\end{equation}
where $\sigma_n$ is the normal electron fluid conductivity. Such a term provides a natural way of including increased dissipation in the pumped state discussed below as a result of the presence of photo-excited carriers. In the equilibrium case, dissipation acts to smooth out sharp features in reflectivity such as the plasma edges.

\subsection{Floquet reflectivity}
\label{sec:floq}

\subsubsection{Floquet eigenstates}
Our goal in this section is to introduce a simple model for Floquet materials and discuss special features  in reflectivity that appear in this model close to parametric resonances. In the next section we will demonstrate that features discussed in this section are ubiquitous, and can be found in more accurate models\cite{Sho19,Marios20}. We model the Floquet medium by assuming that the presence of an oscillating field inside the material results in a time periodic refractive index, $n^2_{\rm driven} (t) = n^2(\omega) + \delta n^2_{\rm drive}  \cos\left( \Omega_d \right)  $. The equations of motion in frequency space for the electric field in the presence of the time-dependent perturbation becomes:
\begin{equation}
\begin{split}     
&\left( k^2 - \frac{ \omega^2 n^2(\omega) }{c^2} \right) E(\omega, k) +\\ &A_{\it drive} \left( E(\omega - \Omega_d, k ) + E(\omega + \Omega_d, k ) \right) = 0,
\label{eq:floqEOM}
\end{split}
\end{equation}
where $A_{\it drive}$ is the mode coupling strength related to the amplitude of the time-dependent drive, which in this section we assume to be constant although, in principle, it may be frequency dependent (see e.g. section \ref{sec:SiC}). Generally, equations of the type
(\ref{eq:floqEOM}) should be solved simultaneously for many frequency components that differ by integer multiples of the drive frequency.
However, to capture parametric resonances in the spectrum, it is sufficient to limit analysis to mixing between only two modes, which are commonly referred to as the signal and idler modes \cite{Ananda16}. The signal frequency is  taken to be the frequency of the probe pulse, whereas the idler frequency is chosen from the condition that the sum of the signal and idler frequency is equal to the drive frequency. There may be other resonant Floquet conditions, such as $\omega_1-\omega_2= \Omega_d$, which do not correspond to parametric generation of excitations by the drive but instead correspond to resonant re-scattering. We postpone discussion of such cases to subsequent publications. Thus we consider
\begin{equation}
    E(t,z) = \left( E_s e^{- i \omega_s t } + E_{id.}^* e^{ + i \omega_{\rm id.} t } \right) e^{i k z}.
\label{eq:ansatz}
\end{equation}
Truncating the eigenmode ansatz to only signal and idler components corresponds to using Floquet degenerate perturbation theory approximation\cite{Eckardt15}. The inclusion of higher harmonic contributions will give rise to sub-leading perturbative corrections. With the ansatz in equation~(\ref{eq:ansatz}), the equations of motion take the form:
\begin{equation}
    \begin{pmatrix} k^2 - k_s^2  &&  A_{\it drive} \\ A_{\it drive} && k^2 - k_{id.}^2\end{pmatrix} \cdot \begin{pmatrix} E_s \\ E^*_{id.} \end{pmatrix} = 0
\end{equation}
where $k_s^2(\omega_s) = \frac{\omega^2_s n^2(\omega_s)}{c^2}$ and $k_{id.}^2(\omega_s) = \frac{\omega_{\rm id.}^2 n^2(\omega_{\rm id.})}{c^2} $ is the momentum of the eigenstate oscillating at the signal frequency or idler frequency respectively in the absence of the parametric drive $A_{\it drive}$. The renormalized eigenvalues are given by:
\begin{subequations}
\begin{align}
    k_{\pm}^2 = \frac{k^2_s + k_{id.}^2}{2} \pm \sqrt{\left(\frac{k_s^2 - k_{id.}^2}{2} \right)^2 + A_{\it drive}^2},
\end{align}
\label{eq:kpm}
\end{subequations}
and the corresponding Floquet eigenstates are:
\begin{subequations}
\begin{align}
    E_{id, \pm }^* &= \alpha_{\pm} E_{s, \pm },\\
    \alpha_{\pm} &= \frac{k_s^2 - k_{id.}^2}{2 A_{\it drive}} \mp \sqrt{ \left(\frac{k_s^2 - k_{id.}^2}{2 A_{\it drive}} \right)^2 + 1}
\end{align}
\label{eq:floqeigen}
\end{subequations}

\subsubsection{Floquet-Fresnel equations}
The eigenstates in equation~(\ref{eq:floqeigen}) represent two transmission channels for the case where the Floquet material is probed at the signal frequency, $E_{\pm}(t,z) = t_{\pm} E_0 \left( e^{- i \omega_s t } + \alpha_{\pm} e^{ + i \omega_{\rm id.} t } \right) e^{i k_{\pm} z}$. Similarly, the transmitted magnetic field is given by $B_{\pm}(t,z) =  k_{\pm} t_{\pm} E_0 \left( \frac{1}{\omega_s} e^{- i \omega_s t } - \frac{\alpha_{\pm}}{\omega_{\rm id.}} e^{ + i \omega_{\rm id.} t } \right) e^{i k_{\pm} z} $. To find the reflectivity, we need to satisfy boundary conditions corresponding to matching of magnetic and electric fields across the boundary oscillating at the signal and idler frequency separately:
\begin{subequations}
\begin{align}
    1 + r_s &= t_+ + t_-, \\
    1 - r_s &= \frac{c k_+}{\omega_s} t_+ + \frac{c k_-}{\omega_s} t_-, \\
    r_{id.} &= \alpha_+ t_+ + \alpha_- t_-,\\
    r_{id.} &= \frac{c k_+}{\omega_{\rm id.}}\alpha_+ t_+ +\frac{c k_- }{\omega_{\rm id.}} \alpha_- t_-
\end{align}
\label{eq:fresnelFloq}
\end{subequations}
where $r_{id.}$ is the coefficient of the light reflected at the idler frequency. The Fresnel-Floquet problem in equation~(\ref{eq:fresnelFloq}) together with equations~(\ref{eq:kpm}) and (\ref{eq:floqeigen}) form a closed set of equations that can be solved to determine the reflectivity $R = |r_s|^2$.

\subsubsection{Perturbation theory for large dissipation }
In order to elucidate the physics of photo-induced resonances, it is instructive to work perturbatively in the parametric driving strength, away from parametric resonance. Since $k_s = k_{id.}$ corresponds to the parametric resonance condition, the small parameter is chosen to be $\xi = \frac{2 A_{\it drive}}{k_s^2 - k_{id.}^2}$. In the limit of small $\xi$, the two solutions can be safely separated into a mostly signal solution and a mostly idler solution. These correspond to expansions of $k_{\pm}^2$ to linear order in $\xi $
\begin{subequations}
\begin{align}
    \tilde{k}_s^2 &\approx k_s^2 + \frac{A_{\it drive} \xi}{2} + \mathcal{O}(A_{\it drive} \xi^3), \\
    \tilde{k}_{id.}^2 &\approx k_{id.}^2 - \frac{A_{\it drive} \xi}{2} + \mathcal{O}(A_{\it drive} \xi^3)
\end{align}
\label{eq:Kpert}
\end{subequations}
where $\tilde{k}_s$ and $\tilde{k}_{id.}$ are the renormalized momenta. The corresponding transmission channels are given by expanding $\alpha_{\pm}$ to leading order in $\xi$:
\begin{subequations}
\begin{align}    
    E_1 =& t_s E_0 \left( e^{- i \omega_s t } - \left( \frac{\xi}{2} + \mathcal{O}(\xi^3) \right) e^{ + i \omega_{\rm id.} t } \right) e^{i \tilde{k}_{s} z}\\
    E_2 =& t_{id.} E_0 \left( \left( \frac{\xi}{2} + \mathcal{O}(\xi^3)\right) e^{- i \omega_s t } + e^{ + i \omega_{\rm id.} t } \right) e^{i \tilde{k}_{id.} z}
\end{align}
\label{eq:Epert}
\end{subequations}
where the eigenmodes have been rescaled in perturbation theory in order to interpret $E_1$ as the channel oscillating primarily at the signal frequency with a perturbative mixing of the term oscillating at the idler frequency, while $E_2$ is the channel oscillating primarily at the idler frequency with a perturbative mixing of a term oscillating at the signal frequency. By integrating out the idler transmission channel the Floquet-Fresnel equations can be reformulated through an effective renormalized refractive index (see Appendix~\ref{app:floqN} for details):
\begin{subequations}
\begin{align}
1 + r_s =& t_s ,\\
1 - r_s =& t_s \tilde{n} 
\end{align}
\end{subequations}
where $\tilde{n}$ is given by:
\begin{equation}
    \tilde{n} = n_{eq.} \left(1 + \frac{A_{\it drive} \xi}{4 k_s^2} + \frac{\xi^2}{4}\frac{ c \tilde{k}_s- \omega_{\rm id.} }{c \tilde{k}_{id.}-\omega_{\rm id.}} \left(  \frac{\tilde{k}_{id.}}{ \tilde{k}_{s}} - 1 \right) \right)
\label{eq:FloqN}
\end{equation}
where $n_{\rm eq.}$ is the equilibrium refractive index. Unlike equilibrium, the dressed Floquet refractive index is allowed to be negative giving rising to parametric amplification of the reflected signal. Equation~(\ref{eq:FloqN}) has two perturbative corrections to second order in the mode coupling strength's amplitude, $A_{\it drive}$: one of order $\xi$ and the other of order $\xi^2$. The term linear in $\xi$ comes from the renormalization of the transmitted wave-vector $\tilde{k}_s$, while the quadratic term results from integrating out the idler channel and therefore originates from interference effects between signal and idler mode. 

\begin{figure}
    \centering
    \includegraphics[trim= 4cm 12cm 0cm 7cm, clip, scale = 0.55]{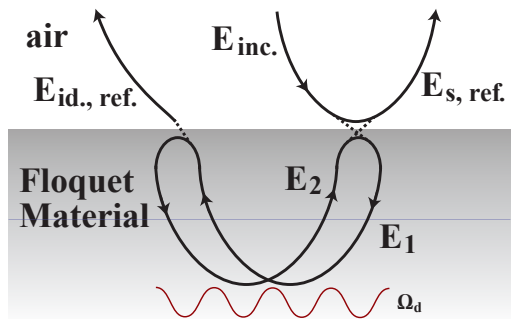}
    \caption{Schematic depiction of optical reflection from a Floquet material. Inside the Floquet material signal and idler frequency components are mixed giving rise to two transmission channels, $E_1$ and $E_2$ propagating in opposite directions. Since the idler component is counter-oscillating compared to the signal, the Floquet drive which mixes signal and idler effectively acts as a wall reflecting each transmission channel and changing its propagation direction. The two channels are coupled at the interface through the boundary conditions. The picture explains schematically the physical mechanism of parametric amplification where the transmission channels are directed towards the surface by the drive.}
\label{fig:floqRefl}
\end{figure}

On parametric resonance within the same band, the phase matching condition between signal and idler $| \mbox{Re} \left(k_s \right) | = | \mbox{Re} \left( k_{\rm id.} \right)|$, implies $\omega_s = \omega_{\rm id.} = \frac{\Omega_d}{2}$, while the sign of each wave-vector is fixed by causality as we show below. The perturbation theory developed above is valid even on resonance provided that the dissipation is high enough. To show this we expand around the parametrically resonant frequency, $\omega_s = \omega_{para}$, with a finite dissipation that we include in a causal way through the substitution $\omega_s \rightarrow \omega_s + i \gamma $. The expressions for the signal and idler wave-vectors are then given by:
\begin{subequations}
\begin{align}
    k_s =& k_{s,0}' + \frac{ \omega_s-\omega_{para} + i \gamma}{v_g(\omega_{para})}, \\
    k_{id.} =& - k_{s,0}' + \frac{ \omega_s-\omega_{para} + i \gamma}{v_g(\omega_{para})}.
\end{align}
\label{eq:ksExpand}
\end{subequations}
In equation~(\ref{eq:ksExpand}),  $v_g(\omega_{para})$ is the group velocity on parametric resonance, and $k_{s,0}'$ the real part of the $k_s$ wave-vector on parametric resonance. Boundary conditions require that the transmitted light vanishes at large distances inside the material or equivalently ${\rm Im}\{k\} >0 $. As a result, the real part of $k_{id.}$ is negative and counter propagates with respect to the mostly signal transmission channel inside the material. This situation is shown schematically in Fig~\ref{fig:floqRefl}. 
Using equation~(\ref{eq:ksExpand}), the parameter $\xi$ is given by
\begin{equation}
    \xi \approx \frac{A_{\it drive} }{2 \left( \omega_s - \omega_{para} + i \gamma \right) \frac{k'_{s,0}}{v_g(\omega_{para}) }}
\end{equation}
and has a lorentzian peak structure. The resonant form of $\xi$ is responsible for resonances in the reflectivity.
In the case of multiple bands, the above result can be generalized by considering that the phase matching condition between signal and idler wave can occur at different signal and idler frequencies $\omega_{para., 1} + \omega_{para,2}  =  \Omega_d$. In this situation we have:
\begin{subequations}
\begin{align}
    k_s =& k_{s,0}' + \frac{ \omega_s -\omega_{para,1} + i \gamma(\omega_{\rm para,1} )}{v_{g}(\omega_{\rm para,1} )}, \\
    k_{id.} =& - k_{s,0}' + \frac{ \omega_{\rm id.} -\omega_{para,1} + i \gamma(\omega_{\rm para, 2})}{v_{g}(\omega_{\rm para,2} )},
\end{align}
\end{subequations}
where the group velocity and dissipation can be different for the different bands at $\omega_{para,1}$ and $\omega_{para,2}$. However, the perturbative parameter $\xi$ for multiple bands takes a form similar to the single band case: 
\begin{equation}
    \xi = \frac{2 A_{\it drive}}{k_s^2 - (k_{id.})^2 } \approx \frac{ A_{\it drive}}{2 \left( \omega_s - \omega_{para,1} + i \gamma_{eff} \right)\times\frac{k_{s,0}'}{v_{eff}  }}
\label{eq:twobands}
\end{equation}
where $ 2v_{eff}^{-1} = v_{g}^{-1}(\omega_{\rm para,1}) + v_{g}^{-1}(\omega_{\rm para,2}) $ and $\gamma_{eff.}  = \frac{v_{eff.}}{2} \cdot \left( \frac{\gamma(\omega_{\rm para,1})}{v_{g}(\omega_{\rm para,1})} + \frac{\gamma(\omega_{\rm para,2})}{v_{g}(\omega_{\rm para,2})} \right)$. Equation~(\ref{eq:twobands}) demonstrates that for small driving strength the resonant behaviour of parametric driving within the same band and in between different bands is the same.

\subsection{Floquet Fresnel Phase Diagram} 
\label{sec:phaseDiag}
Based on our analysis, resonant features in the reflectivity can be classified into four regimes as shown in Fig.~\ref{fig:phase}. Regimes I and II, are in the stable region where dissipation is stronger than the parametric drive. For these cases we can obtain analytic expressions for the changes in reflectivity. To second order in $A_{\it drive}$ we find two contributions to the refractive index given in equation~(\ref{eq:FloqN}). A linear term in $\xi$ arising from band renormalization due to the presence of the drive and another of order $\xi^2$ which appears after integrating out the idler transmission channel and therefore corresponds to interference phenomena between signal and idler transmission. Their relative strength is given by $\frac{\delta n_{linear}}{\delta n_{quadratic}} \propto \frac{\gamma}{v_g k_s}$ on parametric resonance, $|Re(k_s)| = |Re(k_{\rm id.})| $, c.f. equation~(\ref{eq:FloqN}). Interference phenomena dominate, for underdamped photon modes for which $\gamma < v_g k_s$ while for overdamped modes interference phenomena are suppressed and band renormalization is dominant. The corresponding changes to reflectivity are calculated by expanding the reflectivity to linear order in $\delta n$

\begin{subequations}
\begin{align}
    \tilde{n} =& n_{eq.} + \delta n, \\
    \tilde{r}_s =& \frac{1 - \tilde{n}}{1 + \tilde{n}} \approx r_{s,eq.} - \frac{2 \delta n}{ (n+1)^2}, \\
    \tilde{R}_s \approx& R_{s,eq.} - 4 \mbox{Re} \left\{ \frac{\delta n}{ (n+1)^2} r_s^*\right\}.
\end{align}
\label{eq:reflDressed}
\end{subequations}

\textbf{Regime I:} for the usual case of underdamped modes and a single band we can take $r_{s,eq}^*$ and $n_{eq.}$ to be real. Moreover the constant, $ A = \frac{ c \tilde{k}_s- \omega_{\rm id.} }{-c \tilde{k}_{id.} + \omega_{\rm id.}} \left(1 +   \frac{ - \tilde{k}_{id.}}{\tilde{k}_{s}} \right) $, can be expanded around parametric resonance to give : $A = 2  \frac{ n - 1}{n + 1} = 2 r_s $. Under these assumptions, interference of signal and idler gives rise to a double lorentzian dip in reflectivtiy and is reminiscent of EIT. 
\begin{equation}
    \tilde{R}_s \approx R_{s,eq.} \left(1 - \frac{2}{(n + 1)^2} \mbox{Re} \left\{ \frac{C}{\left(\omega_s - \omega_{s,para} + i \gamma \right)^2 } \right\} \right)
\label{eq:underdamped}
\end{equation}
where $C = \frac{v_g^2 A_{\it drive}^2}{4 k_s^2}$ is a constant proportional to the driving intensity.  

\textbf{Regime II: }In the opposite limit of overdamped dynamics in a single band the dominant term comes from the linear in $\xi$ term and the reflectivity takes the form:
\begin{equation}
    \tilde{R}_s \approx R_{s,eq.} + C' \mbox{Re} \left\{ e^{i \theta} \frac{1}{\omega_s - \omega_{s,para} + i\gamma} \right\} 
\label{eq:overdamped}
\end{equation}
where $C' e^{ i \theta} =  - \frac{1}{(n+1)^2} \frac{A_{\it drive}^2 v_g c }{4 k_s^3} $. This feature appears as a plasma edge induced by the drive from a featureless overdamped background as reported in Ref.~\cite{vonHoegen19}.

\textbf{Regime III and IV: } These regimes are not perturbative, however in many cases we can use our simple theory of parametric resonance between two bands to capture the reflectivity of real experiments, by solving equations~(\ref{eq:fresnelFloq}) (a) - (d). In particular, regime IV corresponds to a lasing instability regime where we expect a strong peak in the reflectivity due to parametric amplification and can even be a discontinuous function (as it was also shown in Ref.~\cite{Sho19}. Regime III is an intermediate region where on resonance there is amplification. However away from resonance perturbation theory still holds giving rise to an interesting double dip structure.

\section{Examples of manifestations of parametric resonance in reflectivity} 
\label{sec:parad}
In the previous section, we investigated general aspects of Floquet resonances while being agnostic about microscopic details of the system. In this section, we discuss toy models of realistic dispersion. Pump-induced features in these toy models in the different regimes of the pump-induced phase diagram can be used to build intuition for more complicated multi-band dispersions.

\subsection{Driven plasmon band}
\label{SectionPlasmaBand}

The simplest case of an optical system that we discuss is a single plasmon band, which describes electrodynamics of metals and superconductors. The equilibrium reflectivity in such systems was discussed in Subsec.~\ref{sec:equil}. Maxwell's equations in a superconductor can be written as\cite{Tinkham04}: 
\begin{equation}
   \left(  \omega^2 - \omega_{pl.}^2 + i \frac{\sigma_n}{\epsilon_0}\omega - c^2 k^2 \right) E = 0
\label{EquationPlasmon}
\end{equation}
where $\sigma_n$ represents  the ohmic part of the conductivity and provides dissipation, while the photon obtains a mass given by the plasma frequency. At $\omega = 0$ equation (\ref{EquationPlasmon}) can be solved with $k = i \omega_p$, which corresponds to the skin effect in metals and for superconductors can also be understood as the Meissner effect. The dissipation term with $\sigma_n$ in  (\ref{EquationPlasmon})  can be present in superconductors due to quasiparticles \cite{Bulaevskii94}. The above equations of motion can be represented by the complex refractive index: 
\begin{equation}
    n_{SC} (\omega) = \frac{\omega^2 - \omega_{pl}^2}{\omega^2} + \frac{ i \sigma_{n}}{\epsilon_0 \omega} .
\end{equation}
We model the Floquet material as a system with time-periodic modulation of the plasma frequency $\omega_{pl}$ at  frequency $\Omega_d$. We assume that the modulation frequency is higher than twice the frequency of the bottom of the plasmon band, so that the drive can result in resonant generation of  plasmon pairs. Taking the amplitude of modulation to be  $A_{\it drive} $ we obtain equation~\ref{eq:floqEOM} introduced previously. 

\begin{figure}
    \centering
    \includegraphics[trim= 1.5cm 4cm 1cm 1cm, clip, scale = 0.65]{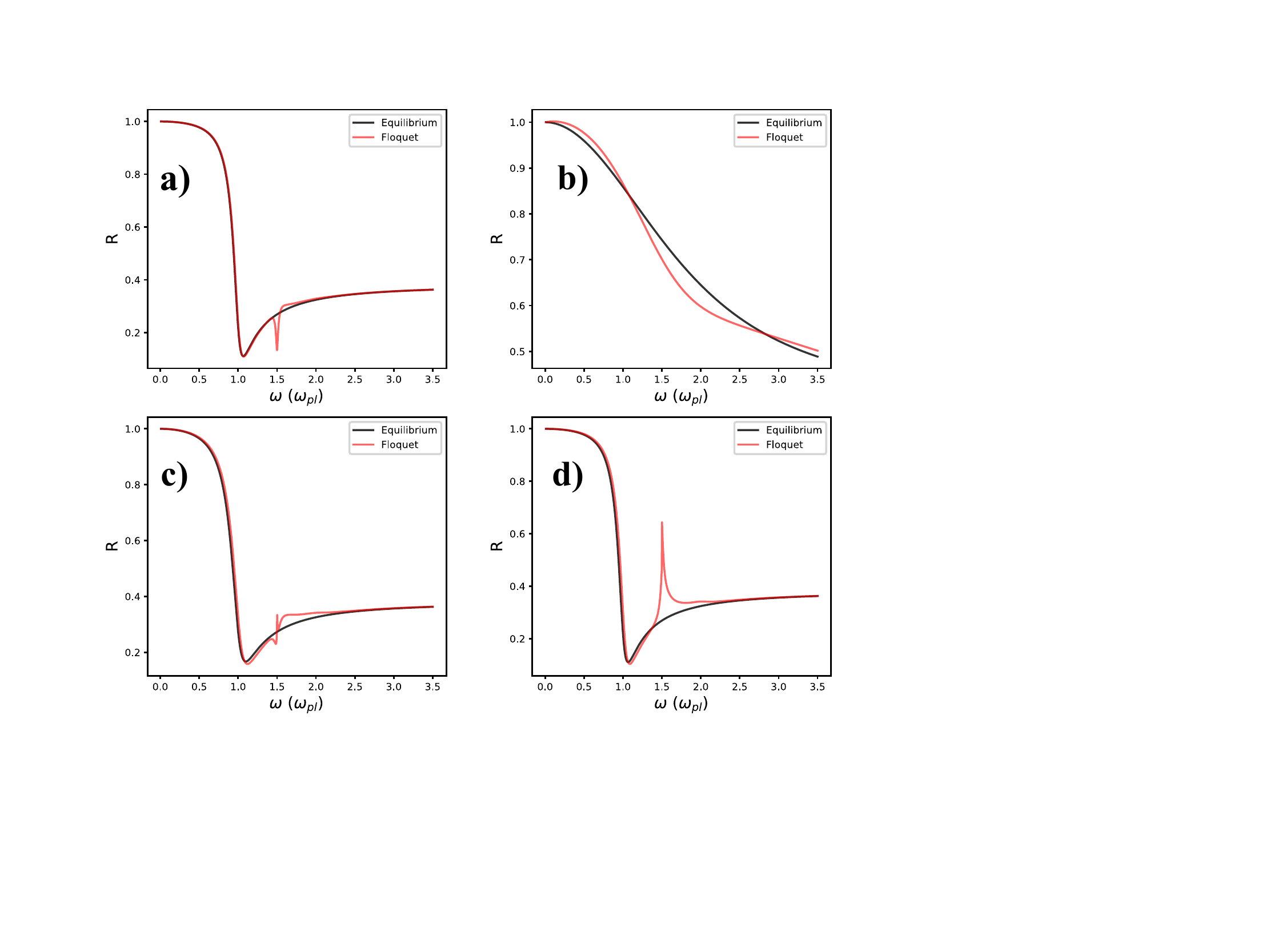}
    \caption{Reflectivity spectra of a plasmon band driven at $\Omega_d = 3 \omega_{pl}$ in the four different regimes of the Phase diagram of Fig.~\ref{fig:phase}. a) Regime I: $\frac{\sigma_n}{\epsilon_0} = 0.064 \omega_{pl}$, $A_{ampl.} = 3 \frac{\omega_{pl}^2}{c^2}$, b) Regime II: $\frac{\sigma_n}{\epsilon_0} = 2 \omega_{pl}$, $A_{ampl.} = 60 \frac{\omega_{pl}^2}{c^2}$, c) Regime III: $\frac{\sigma_n}{\epsilon_0} = 0.1 \omega_{pl}$, $A_{ampl.} = 6 \frac{\omega_{pl}^2}{c^2}$, d) Regime IV: $\frac{\sigma_n}{\epsilon_0} = 0.064 \omega_{pl}$, $A_{ampl.} = 6 \frac{\omega_{pl}^2}{c^2}$. Notice that dissipation suppresses parametric driving effects and a larger oscillation amplitude is needed to produce an appreciable effect in the reflectivity spectra. Notably, in figure (b) which corresponds to an overdamped system, parametric driving gives rise to an interesting structure from a featureless background with a dip on resonance.  }
    \label{fig:plasma}
\end{figure}

Reflectivity spectra in the different regimes of the parametric driving induced phase diagram are plotted in Fig.~\ref{fig:plasma} by tuning the dissipation through the normal conductivity $\sigma_n$ and the amplitude of periodic oscilations $A_{\it drive} $.
\subsection{Floquet-Fresnel reflectivity in a phonon-polariton system}
\label{sec:SiC}

We now want to demonstrate that the four regimes presented in Fig.~\ref{fig:phase} are universal and not limited to a single optical band model. To this end we consider a system that features two branches of optical excitations: a phonon-polariton system corresponding to light coupling to a single IR-active phonon mode. The Hamiltonian describing such  a model can be written as:
\begin{equation}
    H_{\rm ph} = Z E Q + M \omega_{\rm ph}^2 \frac{Q^2}{2} + \frac{\Pi^2}{2 M },
\label{eq:IRHamil}
\end{equation}
where $Q$ is the phonon coordinate, $\Pi$ is the momentum conjugate to $Q$, $\omega_{\rm ph.}$ is the transverse phonon frequency, $M$ is the ion mass, and $Z$ is the effective charge of the phonon mode. 

Solving the equations of motion corresponding to (\ref{eq:IRHamil}) together with Maxwell's equations we obtain two hybrid light-matter modes, corresponding to  the upper and lower polaritons. 
In equilibrium the dispersion and typical reflectivity is given by Fig.~\ref{fig:equil} (b) - (d). The dispersion is modeled by taking a refractive index of the type  (see discussion in section~\ref{sec:equil}) :
\begin{equation}
n(\omega)^2 = \epsilon_{\infty} \left( 1 + \frac{\omega_{\rm pl., phonon}^2 }{-\omega^2 - i \gamma \omega + \omega_{ph.}^2} \right).
\label{eq:refrSiC}
\end{equation}
where in terms of our Hamiltonian parameters the plasma frequency of the phonon is given by, $\omega_{\rm pl. , phonon}^2 = \frac{Z^2}{\epsilon_0 M}$.
The bottom of  the lower polariton branch is at frequency $  \omega_L = \sqrt{\omega_{\rm ph}^2 + \omega_{\rm pl, phonon}^2}$.

\begin{figure}
    \centering
    \includegraphics[trim= 1.5cm 4cm 7.5cm 1cm, clip, scale = 0.6]{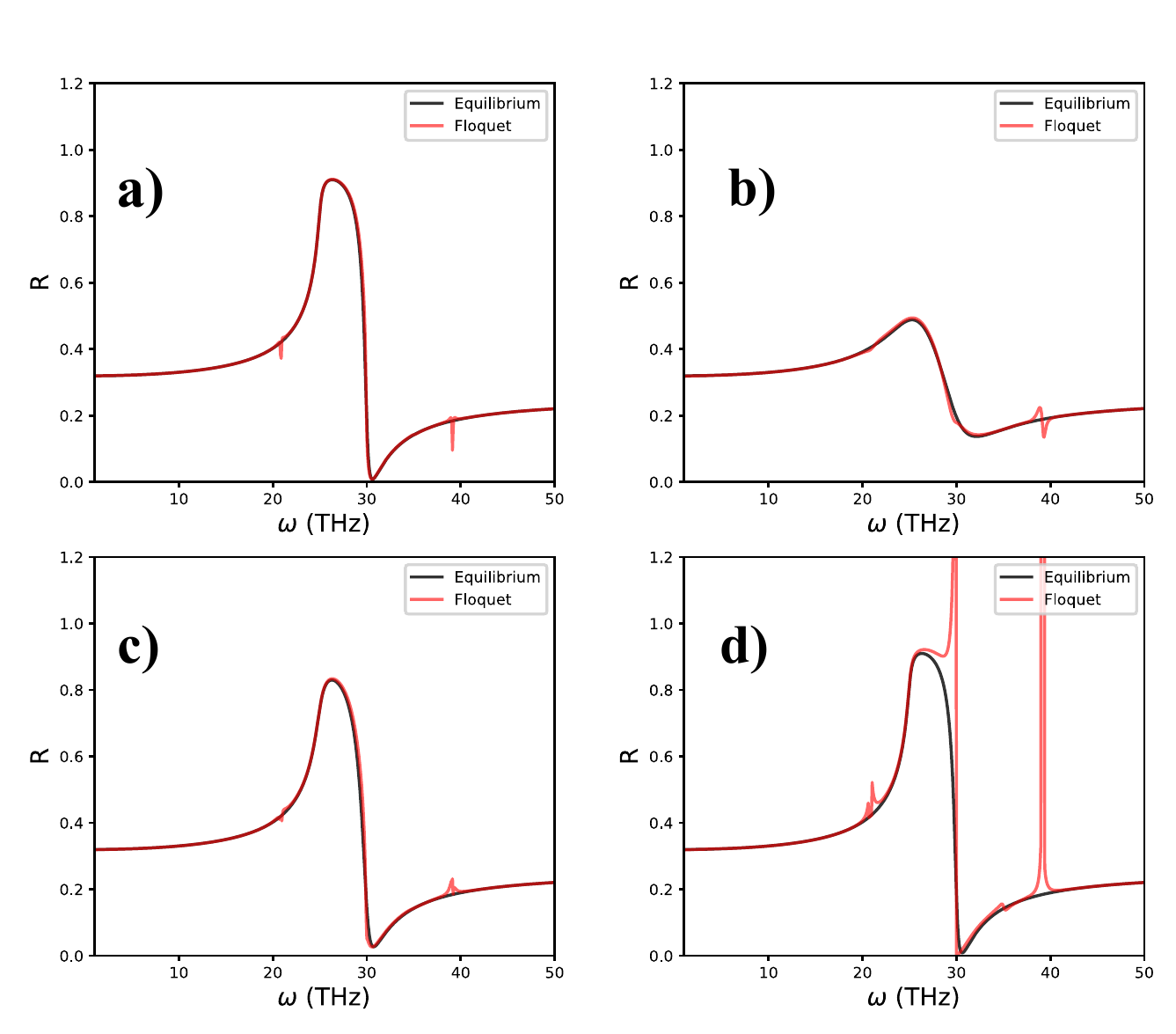}
    \caption{Reflectivity spectra of a phonon-polariton system driven by an effective drive $\Omega_d = 60$ THz through exciting the upper phonon-polariton at $30$ THz. The four regimes of the phase diagram are presented for different parameters of the phonon dissipation and oscillation amplitude. Numbers were chosen such that the transverse phonon is at $\omega_T = 25$ THz and the longitudinal phonon is at $\omega_L = 30$ THz.  a) Regime I: $\gamma = 0.5$ THz , $B = 2.7 \times 10^7 \frac{{\rm THz}^4}{c^2}$, b) Regime II: $\gamma = 5 THz$, $B = 9 \times 10^{7} \frac{{\rm THz}^2}{c^2} $, c) Regime III: $\gamma = 1$ THz , $B = 7000^2 \frac{{\rm THz} ^4}{c^2} $, 
    d) Regime IV: $\gamma = 0.5$ THz,  $B = 8.1\times 10^{7} \frac{{\rm THz}^4}{c^2}$. In regime IV, apart from the expected parametrically resonant instabilities, we find a Fano type feature associated with divergences in the the strength of the phonon mediated parametric drive. This occurs at $\Omega_d - \omega_T  = 35$ THz.   }
    \label{fig:SiC}
\end{figure}

A new feature of the two band system is the possibility of inter-band parametric resonances. The simplest type of optical pump corresponds to resonantly exciting the upper polariton branch at $k=0$, which then results in the parametric drive of the system as frequency $\Omega_d = 2 \omega_L$ \cite{Cartella2018} (for details see Appendix~\ref{app:IRphonon}). This is the situation that we will primarily focus on in this section. As shown in figure \ref{fig:equil}b), in this case one finds a resonant process in which the drive produces one lower and one upper polariton at finite momentum. This process satisfies both momentum and energy are conservation. This resonance leads to singularities in the reflectivity shown in Figure \ref{fig:SiC} at 20 and 40 THz. Another case of parametric resonance corresponds to the drive creating two upper polaritons at zero momentum. This leads to the singularity at $\omega_L=30$ THz in figure \ref{fig:SiC}d).

Another small peak in Figure \ref{fig:SiC}d) (strong drive regime) can be seen at the  frequency of 35 THz. This feature arises from the singularity of the matrix element that mixes the signal and idler frequency components that we pointed out in section \ref{SectionOverview}. In Appendix~\ref{app:IRphonon}, we consider non-linearities in the phonon system of the type
\begin{equation}
   H_{non-linear} = u Q^4.
\label{eq:NonlinHamil}
\end{equation}
and demonstrate that the matrix element $A_{\it drive}$, introduced in equation (\ref{eq:floqEOM}), can be writen as
\begin{equation}
\begin{split}
    &A_{\it drive}(\omega_s) = A_{\it drive,0} + \\&\frac{B}{\left( \omega_s^2 + i\gamma \omega_s - \omega_{ph.}^2 \right)\left(\omega_{\rm id.}^2 +i \gamma\omega_{\rm id.}-\omega_{ph.}^2 \right)}   . 
\label{eq:Amplres}
\end{split}
\end{equation}
The last equation shows that Floquet mixing is dramatically enhanced when either the signal or the idler frequencies are equal to $\omega_{ph}$.
It is also useful to present this result in terms of the effective change of the index of refraction (at the signal frequency) after integrating out contribution of the idler component (see equation (\ref{eq:FloqN})). In the phonon-polariton case we obtain correction to the index of refraction 
\begin{equation}
    \delta n_{\rm phonon} \propto \frac{1}{ \left(\omega_{\rm id.}^2  + i \gamma \omega_{\rm id.} - \omega_{\rm ph.}^2\right)\left( \omega_s^2 + i\gamma \omega_s - \omega_{ph.}^2 \right) },
    \label{delta_n_SiC}
\end{equation}
which shows resonant enhancement around $\omega_s = \omega_{\rm ph.} $ and $\omega_s = \Omega_d - \omega_{ \rm ph.}$. We remind the readers, however, that equation (\ref{delta_n_SiC}) is based on the perturbative treatment of the signal-idler mixing and is not quantitatively accurate in the vicinity of singularities in the reflection coefficient.



\section{Blue shifted edge in bilayer High Tc cuprate $\rm{YBa_2Cu_3O_{6.5}}$}
\label{sec:exp}

An experimental realization of the driven single plasmon edge comes from terahertz pump and probe experiments in $\rm{YBa_2Cu_3O_{6.5}}$\cite{Liu20}. In equilibrium, $\rm{YBa_2Cu_3O_{6.5}}$ is a bi-layer superconductor with a Josephson plasmon at $0.9$ THz. The low energy optical response for light polarized along the c-axis of the material is captured by the equations of motion\cite{Tinkham04}:
\begin{equation}
     \left( n^2_0\left( \omega^2 - \omega_{\rm pl.}^2 \right) +i \frac{\sigma_n}{\epsilon_0} \omega - c^2 k^2  \right) E = 0
\end{equation}
where $\sigma_n$ represents the conductivity of the normal state electron fluid which provides dissipation for the Josephson plasmon, $\omega_{pl}$ the Josephson plasma frequency and $n_0$ the static refractive index inside the material. This photon dispersion is shown in Fig.~\ref{fig:equil}(a) with a gap $\omega_{JP} \sim 0.9$ THz leading to a Josephson plasma edge at that frequency in the equilibrium optical reflectivity. Equivalently, the equations of motion can be represented by the refractive index:
\begin{equation}
    n_{SC}(\omega)  = n^2_0 \left( \frac{ \omega^2 - \omega_{\rm pl.}^2}{\omega^2} + i\frac{\sigma_n}{\epsilon_0 n^2_0 \omega} \right),
\end{equation}
substituted in equation~(\ref{eq:eomEquil}).

\begin{figure}
    \centering
    \includegraphics[trim= 1.5cm 6cm 2cm 1cm, clip, scale = 0.45]{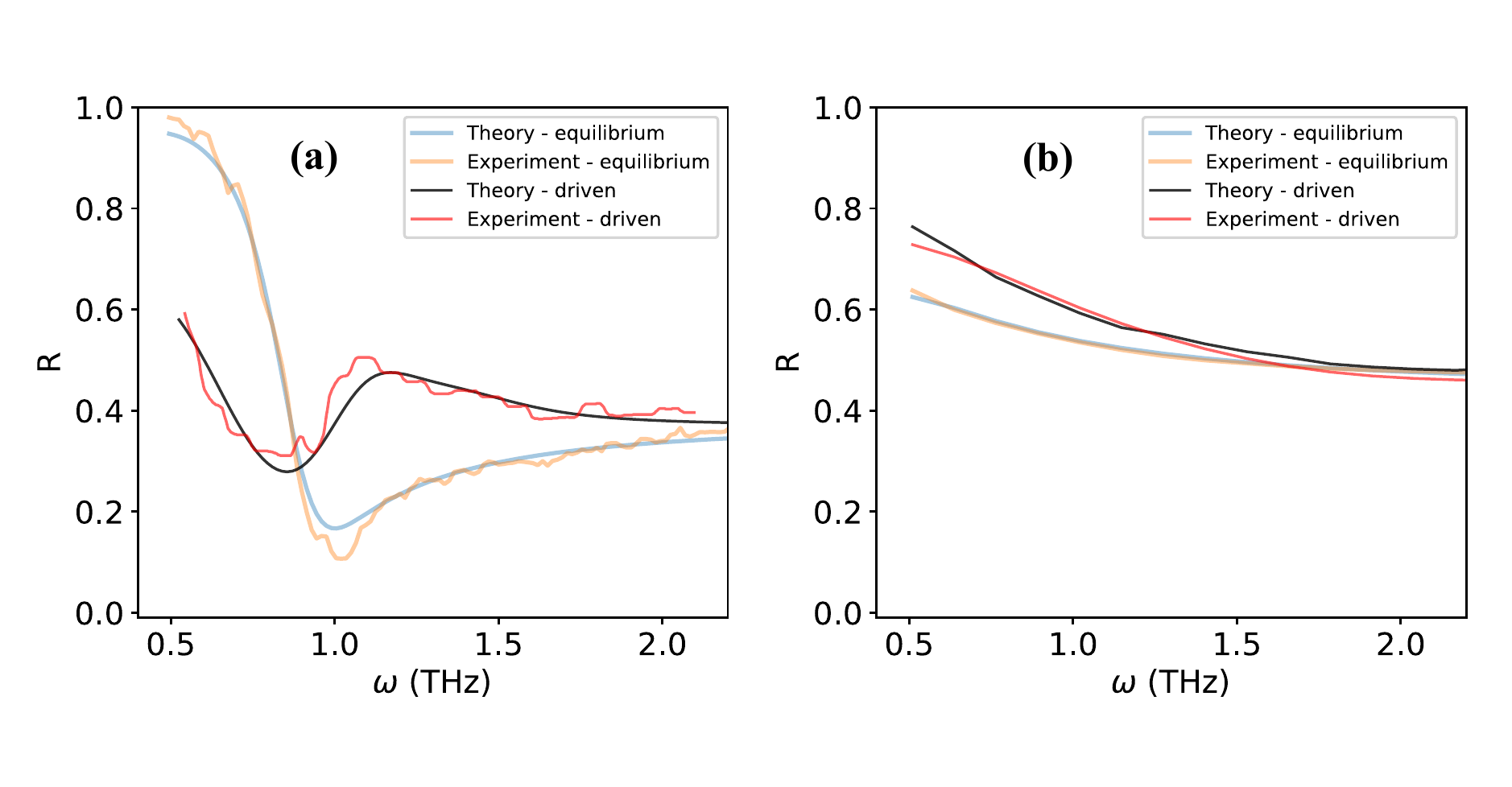}
    \caption{Optical reflectivity spectra of $\rm{YBa_2Cu_3O_{6.5}}$ extracted from Ref.~\cite{Liu20}, re-plotted with the permission of the authors and fitted with the theory presented in this paper. The photo-induced reflectivity edge is well captured by our simple model and suggests that Josephson plasmons are parametrically driven by a coherently oscillating mode. (a) Reflectivity spectra at $T = 10 K$ (below $\rm T_c$) shows a dip peak structure around 1 THz corresponding to regime (I) of our phase diagram. (b) Reflectivity spectra at $T = 100 K$ (above $\rm T_c$), is fitted with our theory assuming an overdamped Josephson plasmon edge in the pseudogap regime. Parametric driving produces changes in reflectivity consistent to regime (II) of our phase diagram. Fitting parameters reported in Appendix~\ref{app:YBCO}.
    }
    \label{fig:YBCO}
\end{figure}

We use our model to fit experimental data presented in reference\cite{Liu20} (reprinted here with the author's permission). Parameters used in this section to produce the figures are tabulated in Appendix~\ref{app:YBCO}. We consider first a low temperature state in the superconducting regime, $T = 10 K $, and model pumping as parametrically driving Josephson plasmons\cite{Marios20, vonHoegen19}. Using our simple model, we find excellent agreement with experiments shown in Fig.~\ref{fig:YBCO}, and interpret the edge at $\sim 1$ THz to be a consequence of parametric resonance from a drive at $\sim 2 $ THz and corresponds to the intermediate Regime IV in the phase diagram. To fit the data, we need to assume that the normal state conductivity, $\sigma_n$ is increased in the pumped state by photo-excited quasiparticles, but also $\omega_{\rm pl.}^2$ which is proportional to the superfluid density is decreased. Remarkably, our simulation shows that even if we assume a suppressed superfluid density, we still find a blue-shifted edge as a result of internal oscillating fields parametrically driving Josephson plasmons. To fit the photo-induced edge above $T_c$, we model the pseudogap phase as a superconductor with overdamped dynamics and a reduced plasma resonance frequency. In Fig.~\ref{fig:YBCO} (b) we are able to fit the data assuming parametric driving at the same frequency as the low temperature data. Our theory suggests that reflectivity data from pump and probe experiments in the pseudogap phase of $\rm{YBa_2Cu_3O_{6.5}}$ correspond to Regime II of our phase diagram, which shows that a photo-induced edge appears as a result of parametric driving of overdamped photon modes. 


\section{Discussion and outlook}
\label{sec:concl}

In this paper we developed a theory that allows to compute optical reflectivity of materials with oscillating collective modes. We demonstrated that using only a few phenomenological coefficients, which parametrize the frequency dependent refractive index, as well as the frequency of the oscillations driving the system, it is possible to predict the position of the photo-induced resonances associated with parametric resonances. To obtain the shape of the resonant feature, one also needs to include information about the amplitude of the drive and dissipation of collective modes. In particular, we found that when dissipation dominates over parametric drive the system develops a lorentzian shaped dip, which arises from the interference of signal and idler transmission channels. At stronger drives the dip turns into a peak and reflectivity can exceed one, indicating parametric amplification of the probe pulse. We also discussed interesting double dip crossover behaviour between the overdamped and amplification regimes. Our results should be ubiquitous in strongly driven systems where the excitation of a well-defined collective mode can act as the external periodic drive. 




Our analysis demonstrates that parametric resonances provide a general universality class of reflectivity features from which both dynamical and static properties of the system can be extracted. This puts them in the same category as previously studied Fano resonance features and EIT\cite{Limonov17}. Despite the simplicity of our model the resulting reflectivity spectrum can be quite rich, as shown in the phase diagram in Fig.~\ref{fig:phase}. Our results provide a tool for analyzing a variety of photo-induced features that have been observed in experiments but have not been given theoretical interpretation until now. We show that photo-induced features, such as a photo-induced edge, can serve as a reporter of a long lived collective mode excited in the material during pumping and a pre-cursor of a lasing instability that can occur in the system at stronger drives. As a concrete case study we analyzed experimental results of the pump-induced changes of reflectivity in a layered superconductor $\rm{YBa_2Cu_3O_{6.5}}$ at frequencies close to the lower Josephson plasmon edge. We find that we can obtain an accurate fit to the experimental data if we include strong renormalization of the equilibrium parameters, such as enhancement of real conductivity due to the photo-excitation of charge carriers during the pump. 

A natural generalization of the above framework is the inclusion of time dependent drives at several frequencies. This is important, for example, for analyzing Floquet drives with finite spectral width or including finite lifetime of collective modes. In this case different oscillating modes are expected to compete with each other, leading to a inhomogenious broadening of the dip / peak features predicted in this work.

\section*{Acknowledgements}
SRUH  and RDA acknowledge support from the DARPA DRINQS program  (Grant No. D18AC00014). DP thanks support by the Israel Science Foundation (grant 1803/18).

\appendix

\section{Derivation of Floquet refractive index in the stable regime}
\label{app:floqN}
In this section we derive the Floquet refractive index shown in equation (\ref{eq:FloqN}). Using equations (\ref{eq:Kpert}) and (\ref{eq:Epert}) we derive the perturbative Floquet-Fresnel equations:
\begin{subequations}
\begin{align}
    1 + r_s &= t_s + \frac{\xi}{2} t_{id.},\\
    1 - r_s &= t_s \frac{c \tilde{k}_{s}}{\omega_s} + \frac{\xi}{2} \frac{c \tilde{k}_{id.}}{\omega_s}, \\
    r_{id.} &= - \frac{\xi}{2} t_s  + t_{id.}, 
    \label{eq:erid}\\
    r_{id.} &= - \frac{\xi}{2} \frac{c \tilde{k}_{s}}{\omega_{\rm id.}} t_{s}  + \frac{c \tilde{k}_{id.}}{\omega_{\rm id.}} t_{id.}
    \label{eq:brid}
\end{align}
\end{subequations}
We can integrate out the effects of the idler channel, by using equations (\ref{eq:erid}),(\ref{eq:brid}): 
We wish to use the boundary conditions oscillating at the idler frequency to solve for $t_{id.}$ in terms of $t_s$:
\begin{equation}
    t_{id.} = \frac{\xi}{2} \frac{ c \tilde{k}_s- \omega_{\rm id.} }{c \tilde{k}_{id.}-\omega_{\rm id.}} t_s
\end{equation}
These lead to the boundary conditions:
\begin{subequations}
\begin{align}
    1 + r_s =& t_s (1 + \frac{\xi^2}{4} \frac{ c \tilde{k}_s- \omega_{\rm id.} }{c \tilde{k}_{id.}-\omega_{\rm id.}}) , \\
    1 - r_s =& t_s \frac{ c \tilde{k}_s}{\omega_s} \left( 1  + \frac{\xi^2}{4} \frac{ c \tilde{k}_s- \omega_{\rm id.} }{c \tilde{k}_{id.}-\omega_{\rm id.}} \frac{\tilde{k}_{id.}}{ \tilde{k}_{s}} \right)
\end{align}
\end{subequations}
After re-scaling the transmission coefficient $t_s$, the above equation can be cast in the familiar form:
\begin{subequations}
\begin{align}
1 + r_s =& t_s ,\\
1 - r_s =& t_s \tilde{n} 
\end{align}
\end{subequations}
allowing us to encode the effects of driving into an effective renormalized refractive index. In fact, the possibility for the dressed refractive index to be negative is what gives rise to phenomena such as parametric amplification of reflectivity. The renormalized refractive index is found to be:
\begin{subequations}
\begin{align}
    \tilde{n} \approx& \frac{ c \tilde{k}_s}{\omega_s} \frac{ 1 + \frac{\xi^2}{4}  \frac{ c \tilde{k}_s- \omega_{\rm id.} }{c \tilde{k}_{id.}-\omega_{\rm id.}} \cdot \frac{\tilde{k}_{id.}}{ \tilde{k}_{s}}}{1 + \frac{\xi^2}{4}  \frac{ c \tilde{k}_s- \omega_{\rm id.} }{c \tilde{k}_{id.}-\omega_{\rm id.}}}, \\
    \tilde{n} \approx& n_{eq.} \left(1 + \frac{A_{\it drive} \xi}{4 k_s^2} + \frac{\xi^2}{4}\frac{ c \tilde{k}_s- \omega_{\rm id.} }{c \tilde{k}_{id.}-\omega_{\rm id.}} \left(  \frac{\tilde{k}_{id.}}{ \tilde{k}_{s}} - 1 \right) \right)
\end{align}
\label{eq:dressed_pert}
\end{subequations}
as reported in equation~(\ref{eq:FloqN}).

\section{IR phonon mediated drive}
\label{app:IRphonon}
The equation of motion for the phonon given by the hamiltonian in equation~(\ref{eq:IRHamil}) and (\ref{eq:NonlinHamil}) is
\begin{equation}
    \left( \partial_t^2 + \gamma \partial_t + \omega_0^2 + 4 u Q^2 \right) Q = Z E.
\end{equation}
Using a gaussian ansatz for the phonons we can linearize the above equation as:
\begin{equation}
     \left( \partial_t^2 + \gamma \partial_t + \omega_{\rm ph.}^2 + 12 u \expe{Q^2} \right) Q = \frac{Z}{M} E.
\label{eq:phononeom}
\end{equation}
The phonon mode appears in the Maxwells equations as:
\begin{equation}
    \left( \frac{1}{c^2}\partial_t^2 - k^2 \right) E = - Z \partial^2_t Q.
\end{equation}
Oscillating collective modes inside the material will affect the above linear system through oscillations of $\expe{Q^2} = \expe{X^2}_0 + A \left( e^{i \Omega_d t} + e^{-i \Omega_d t}  \right)$. Such a term can arise by pumping the system on resonance with the upper polariton, such that $\expe{Q} = A' \cos{ \omega_{L} t} $, where $\omega_L^2 = \omega_{ph.}^2 + \frac{Z^2}{M} $, the frequency of the upper polariton at $k = 0$. Alternatively, for a pumping protocol at high frequencies, the upper polariton fluctuations, $\expe{Q^2} $, can be driven linearly by a Raman process. In both cases, the driving frequency would be twice the upper plasmon frequency $\Omega_d = 2 \omega_L $. However, in general $\Omega_d$ can also correspond to a different frequency not included in our model. Absorbing $\expe{Q^2}_0$ in the definition of $\omega_{\rm ph.} $ and expanding in equation~(\ref{eq:phononeom}) $Q$ in signal and idler components, $Q = Q_s e^{- i \omega_s t} + Q_{\rm id.} e^{i \omega_{\rm id.} t} $ we have
\begin{widetext}
\begin{equation}
\begin{pmatrix} Q_s \\ Q_{id} \end{pmatrix} = \begin{pmatrix} \frac{Z}{ \omega_s^2 + i \gamma \omega_s - \omega_{\rm ph.}^2 } && 0 \\ 0 &&\frac{Z}{ \omega_{\rm id.}^2 +i \gamma \omega_{\rm id.} - \omega_{\rm ph.}^2}  \end{pmatrix} \cdot \begin{pmatrix} E_{s} \\ E_{\rm id}  \end{pmatrix} + \frac{Z A}{\left(\omega^2 + i \gamma \omega - \omega_{\rm ph.}^2 \right)\left(\omega_{\rm id.}^2 +i \gamma \omega_{\rm id.} - \omega_{\rm ph.}^2 \right) } \begin{pmatrix} E_{\rm id} \\ E_s \end{pmatrix} .
\label{eq:Floqphonon}
\end{equation}
\end{widetext}
Substituting equation~\ref{eq:Floqphonon} in Maxwells equation we find the equations of motion for the signal and idler component to be:
\newline
\begin{subequations}
\begin{align}
\left( \frac{n^2_{eq.} (\omega_s) }{c^2} \omega_s^2 - k^2 \right) E_s+  A_{\rm drive, s} (\omega_s, \omega_{\rm id} ) E_{\rm id}=&0 ,\\
\left( \frac{n^2_{eq.} (\omega_{\rm id.}) }{c^2} \omega_s^2 - k^2 \right) E_s+  A_{\rm drive, id} (\omega_s, \omega_{\rm id} ) E_{\rm id}=&0
\end{align}
\end{subequations}
where the signal and idler driving amplitude, $A_{\rm drive, s}$ and $A_{\rm drive, id}$ is given by:
\begin{subequations}
\begin{align}
\begin{split}   
&A_{\rm drive,s} =\\
&\frac{Z^2 A \omega_s^2}{\left(\omega^2 + i \gamma \omega - \omega_{\rm ph.}^2 \right)\left(\omega_{\rm id.}^2 +i \gamma \omega_{\rm id.} - \omega_{\rm ph.}^2 \right) } ,
\end{split} \\
\begin{split} 
     &A_{\rm drive,s} = \\
     &\frac{Z^2 A \omega_{\rm id}^2}{\left(\omega^2 + i \gamma \omega - \omega_{\rm ph.}^2 \right)\left(\omega_{\rm id.}^2 +i \gamma \omega_{\rm id.} - \omega_{\rm ph.}^2 \right) } ,
\end{split}
\end{align}
\end{subequations}
justifying the resonant structure presented in equation~\ref{eq:Amplres}.

\section{Fitting parameters for $\rm{YBa_2Cu_3O_{6.5}}$ data} 
\label{app:YBCO}

As mentioned, equilibrium is modeled via the equations of motion of photons in a superconductor:
\begin{equation}
     \left( \omega^2 + i \frac{\sigma_n}{\epsilon_0} \omega - \left(  \omega_{\rm pl.}^2 + \frac{c^2}{n^2} k^2 \right) \right) E(\omega) = 0.
\end{equation}

Driving is taken into account by mixing signal and idler frequency contributions arising from a periodic drive at $\Omega_d$:
\begin{subequations}
\begin{align}
\begin{split}
&\left( \omega_s^2 + i \frac{\sigma_n}{\epsilon_0} \omega_s - \left(  \omega_{\rm pl.}^2 + \frac{c^2}{n^2} k^2 \right) \right) E(\omega_s,k) +\\
&A_{\it drive} E( - \omega_{\rm id.},k )= 0,
\end{split}\\
\begin{split}
&\left( \omega_{\rm id.}^2 + i \frac{ \sigma_n}{\epsilon_0} \omega_{\rm id.} - \left(  \omega_{\rm pl.}^2 + \frac{c^2}{n^2} k^2 \right) \right) E(- \omega_{\rm id.},k) +\\
&A_{\it drive} E( \omega_{\rm s},k )= 0,
\end{split}
\end{align}
\end{subequations}

To fit the data, we first fit the parameters $ \{ \omega_{pl.}, \sigma_n, n \}$ to the equilibrium reflectivity, and then fit the driving frequency $\Omega_d$ and driving amplitude, $A_{\it drive}$ to match the driven reflectivity spectra. 
\paragraph{Below $T_c$:} The experimental data taken at 10 K, with pumping frequency of $19.2$ THz with a width of $1$ THz and electric field amplitude $1 MV /cm $. The equilibrium is fitted with $ \omega_{pl.} = 0.9$ THz , $\sigma/\epsilon_0  = 2.7$ THz  , $n = 4.2$. In the driven state we use $\Omega_d = 2.1$ THz, $A_{\it drive} = 8.4 {\rm\frac{THz^2}{c^2} } $, $\omega_{pl.} = 0.6$ THz and $\sigma_n/\epsilon_0 = 5.5$ THz. From our fit we predict that dissipation has increased due to pumping but also the interlayer Josephson coupling has decreased during the pump. We see that the edge appears even if the Josephson coupling is suppressed.

\paragraph{Above $T_c$ : } The experimental data taken at 10 K, with pumping frequency of $19.2$ THz with a width of $1$ THz and electric field amplitude $3 MV /cm $. Equilibrium is found to be overdamped with $ \omega_{pl.} = 0.1$ THz , $\sigma/\epsilon_0  = 25.8$ THz  , $n = 5$. The pumped reflectivity is fitted with $\Omega_d = 3.8$ THz, $A_{\it drive} = 64 {\rm\frac{THz^2}{c^2} } $, $\omega_{pl.} = 0.1$ THz and $\sigma_n/\epsilon_0 = 54$ THz.

Finally, both signals where convoluted with a Gaussian broadening function, with standard deviation $0.05$ THz.

\bibliography{references}{}

\end{document}